# Recovering complex ecological dynamics from time series using state-space universal dynamic equations

Jack H. Buckner[*,1], Zechariah D. Meunier[1], Jorge Arroyo-Esquivel[2], Nathan Fitzpatrick[3], Ariel Greiner[4,5], Lisa C. McManus[3], James R. Watson[1]

**Abstract**

Ecological systems often exhibit complex nonlinear dynamics like oscillations, chaos, and regime shifts. Universal dynamic equations have shown promise in modeling complex dynamics by combining known functional forms with neural networks that represent unknown relationships. However, these methods do not yet accommodate the forms of uncertainty common to ecological datasets. To address this limitation, we developed state-space universal dynamic equations by combining universal differential equations with a state-space modeling framework, accounting for uncertainty. We tested this framework on two simulated and two empirical case studies and found that this method can recover nonlinear biological interactions that produce complex behaviors including chaos and regime shifts. Their forecasting performance is context-dependent with the best performance on chaotic and oscillating time series. This new approach leveraging both ecological theory and data-driven machine learning offers a promising new way to make accurate and useful predictions of ecosystem change.

## Introduction

Ecological systems often exhibit complex dynamic phenomena like oscillations, chaos, and nonlinear regime shifts. Quantifying these dynamics is valuable for both scientific understanding and ecosystem management. Identifying thresholds where regime shifts occur can help determine when management interventions are required to maintain desired states or to reverse undesired regime shifts (Suding & Hobbs 2009). Furthermore, models that capture nonlinear ecosystem dynamics can improve the ability to forecast ecosystem states, creating the potential for managers to respond proactively to ecological change (Perretti *et al.* 2013).

Time series data from long-term monitoring programs are a valuable source of information for observing changes in ecological systems over time. From these data we can extract correlative information about the internal and external factors causing changes in the ecosystem, which in turn allows us to quantify the mechanisms driving ecosystem dynamics. However, modeling

---

[*] Corresponding author: Jack H. Buckner (email: bucknerjo@oregonstate.edu)
[1] College of Earth, Ocean, and Atmospheric Sciences, Oregon State University, Corvallis, OR, USA
[2] California Department of Fish and Wildlife, West Sacramento, CA, USA
[3] Hawaiʻi Institute of Marine Biology, University of Hawaiʻi at Mānoa, Kāneʻohe, HI, USA
[4] Department of Biology, University of Oxford, Oxford, UK
[5] Centre for Infectious Disease Dynamics, Pennsylvania State University, University Park, PA, USA

these data requires a flexible modeling framework that can identify nonlinear relationships in the data in the context of noisy observations and stochastic variation in both observed and unobserved factors. Recent advances in machine learning have produced a promising new class of models for this task: universal differential equations (Arroyo-Esquivel *et al.* 2023; Bonnaffé *et al.* 2021; Rackauckas *et al.* 2021) for continuous time models and universal difference equations for discrete time models, collectively called universal dynamic equations (UDEs). UDEs combine specific parametric functions to capture known relationships and physical constraints (i.e., ecological theory) with neural networks to learn unknown relationships directly from data.

The use of artificial neural networks makes these methods especially promising because they can represent arbitrary nonlinear relationships in the dataset and scale in a computationally efficient way for both the number of input dimensions and the size of the dataset (Cheridito *et al.* 2022). UDEs can improve forecasting by incorporating prior information about a system's structure through parametric functions (Arroyo-Esquivel *et al.* 2023). The flexibility of neural networks within UDEs makes them valuable for inferring unknown nonlinear functions such as species interactions directly from time series data (Bonnaffé *et al.* 2021). Furthermore, UDEs have the same mathematical structure as models commonly used for ecological theory, increasing their interpretability when compared to other machine learning techniques such as transformers (Vaswani *et al.* 2023), recurrent neural networks, and long short-term memory networks (Hochreiter & Schmidhuber 1997).

Despite the promise of UDEs for describing nonlinear dynamics, standard formulations of UDEs (e.g., Arroyo-Esquivel *et al.* 2023; Bonnaffé *et al.* 2021) do not account for the combination of stochastic external forcing and noisy observations common in ecological data. In fact, simple methods for training UDEs often fail when confronted with highly variable time series data that include either chaotic or stochastic dynamics (Turan & Jäschke 2022). To address these issues, we developed a new modeling framework called state-space UDEs that embeds UDEs within a state-space modeling framework (*sensu* Auger-Méthé *et al.* 2021).

State-space models describe time series data using two sub-models: a process model that describes changes in the state of the system and a data model that describes the relationship between the state variables of a dynamical system and the measurements in the dataset (Auger-Méthé *et al.* 2021). These two sub-models allow state-space models to simultaneously account for two types of uncertainty: imperfect forecasts of the system's state (i.e., process error) and noisy or imperfect measurements (i.e., observation error). We hypothesize that state-space UDEs will leverage the flexibility of the UDE modeling framework for describing nonlinear local ecosystem dynamics while maintaining the ability of state-space models to accommodate the uncertainty common in ecological data.

We test the ability of the state-space UDE modeling framework to quantify nonlinear dynamics from noisy ecological data using both simulated and empirical examples. The simulated examples allow us to test the model predictions with an unlimited number of replicates when the ground truth is known, while the empirical examples ensure that the models work under the constraints of real-world datasets. To assess model performance under different dynamic regimes, we examine two pairs of systems. The first pair includes a simulated three-species food chain model and empirical fisheries datasets that exhibit cycling. These examples feature intrinsic nonlinear dynamics that produce high temporal variability through oscillation and chaos. For the second pair, we simulate data from a model representing kelp forest and urchin barren dynamics, and use an empirical dataset documenting desertification in an arid rangeland (Christensen *et al.* 2021). In these systems, positive feedback loops interact with external forcing to cause transitions between alternative stable states. Using these examples, we demonstrate the usefulness of state-space UDEs in generating accurate forecasts and revealing underlying mechanisms of ecosystem dynamics from time series data.

**State-space Universal Dynamic Equations**

State-space UDEs describe a sequence of observations $\boldsymbol{y}_t$ from a dynamical system with a sequence of state variables $\boldsymbol{u}_t$ (Fig. 1). These two quantities are linked by the data model $h(\boldsymbol{u}_t, \boldsymbol{y}_t)$, which describes the probability of observing $\boldsymbol{y}_t$ given the true state of the system $\boldsymbol{u}_t$. In our examples, the observations are equal to the states plus a normally-distributed measurement error $\epsilon_t$ with mean zero and variance $\sigma^2$. For multivariate datasets, we assume the measurement error terms are independent and identically distributed. However, in principle, the data model $h(\boldsymbol{u}_t, \boldsymbol{y}_t)$ can accommodate alternative data types by modifying the function linking the observations with the data and its distributional assumptions over the error terms.

A universal dynamic equation (Fig. 1A,B) describes changes in the system's state between observations $\boldsymbol{u}_t \rightarrow \boldsymbol{u}_{t+\Delta t}$ (Fig. 1C). The universal dynamic equation defines a function $F(\boldsymbol{u}_t, \boldsymbol{X}_t, t, \Delta t; \theta_f) \approx \boldsymbol{u}_{t+\Delta t}$, which forecasts the state of the system after a period of $\Delta t$, given the current state $\boldsymbol{u}_t$, covariate factors $\boldsymbol{X}_t$, and parameters estimated from the data $\theta_f$. In the discrete-time case, the function $F$ is defined directly with a combination of known functions and neural networks. In the continuous-time case, we define $F$ by integrating the right-hand side of an ordinary differential equation (ODE) $d\boldsymbol{u}/dt = f(\boldsymbol{u}, \boldsymbol{X}(v); \theta_f)$, where $f$ is defined by a combination of known functions and neural networks from $t$ to $t + \Delta t$:

1) $F(\boldsymbol{u}_t, t, \Delta t; \theta_f) = \boldsymbol{u}_t + \int_t^{t+\Delta t} f(\boldsymbol{u}(v), \boldsymbol{X}(v); \theta_f) dv.$

The function $\boldsymbol{X}(t)$ describes the values of covariates at each point in time. We construct continuous-time covariate functions from discrete-time samples using linear splines. We model the forecasting error as a normally-distributed random variable $\boldsymbol{v}_t$ with mean zero and variance $\Delta t \tau^2$, but like the data model, these distributional assumptions can be relaxed within the state-space UDE framework.

We train the model (Fig. 1D,E) by solving for the unobserved state variables $\hat{\boldsymbol{u}}_t$ and the UDE model parameters $\theta_f$ to maximize the log-likelihood of the data $\boldsymbol{y}_t$ plus a set of priors $P(\theta_f)$ over the model parameters. The form of the log-likelihood function depends on the distribution of the model error term; in the case of a normal distribution, the likelihood is equal to the sum of squared errors weighted by the inverse of the variance terms, plus a constant $c$:

2) $L(\boldsymbol{y}_t | \hat{\boldsymbol{u}}_t, \theta_f) = -\frac{1}{\sigma^2} \sum_{t=1}^{T} (\boldsymbol{y}_t - \hat{\boldsymbol{u}}_t)^2 - \sum_{t=2}^{T} \frac{1}{\Delta t \tau^2} \left( \hat{\boldsymbol{u}}_{t+1} - F(\boldsymbol{u}_t, t, \Delta t; \theta_f) \right)^2 - P(\theta_f) + c.$

We use the priors over the known parametric functions to incorporate additional biological information and normally-distributed priors over the weights of the neural networks to control the smoothness of the functions learned by the neural network.

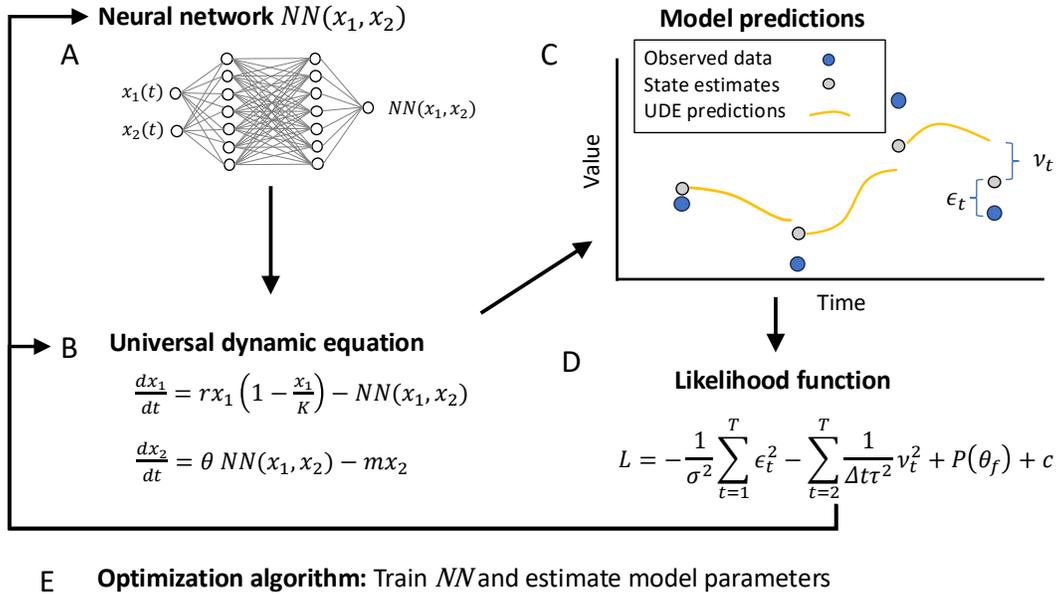

**Figure 1:** Schematic diagram of a state-space universal dynamic equation for a predator-prey model ($x_1$ is prey, $x_2$ is predator). A) Artificial neural networks represent unknown relationships between state variables. B) The neural network output is combined with known functional forms (i.e., Lotka-Volterra predator-prey model with logistic growth of the prey) to build a model of the system. C) The universal differential equation model predicts changes in the state of the system between observations (yellow lines). The state-space model estimates the true but unknown state of the system at each time point (gray dots), which is compared to the observed data (blue dots) to calculate observation errors $\epsilon_t$. The differences between the predictions and the estimated states are process errors $v_t$. D) The log-likelihood function quantifies the performance of the model by computing the weighted sum of squared observation errors and process errors. E) An optimization algorithm trains the neural network and estimates the model parameters and unknown states by maximizing the log-likelihood function.

**Results**

*Recovering chaotic dynamics from noisy time series*

We tested the ability of the state-space UDE framework to recover complex nonlinear dynamics by training a model on simulated time series from a three-species food chain ODE system with chaotic dynamics (Fig. 2A, Hastings & Powell 1991). We trained a nonparametric UDE that uses only a neural network to represent the dynamics (sometimes called a neural ordinary differential equation or NODE; Chen *et al.* 2019, Bonnaffé *et al.* 2021). This model reduces the influence of choices about the model structure on its performance, isolating the influence of the state-space formulation and training routine. We tested the model's performance under process error by adding white noise to the growth rate of the primary producer in the food chain model. We also tested the effect of observation error by adding normally-distributed measurements errors with mean zero and variance $\sigma_\epsilon^2$ to the log abundances of each species. Note that details for all models are provided in the Methods.

The state-space UDE model outperformed four alternative models in one-step ahead predictions when measurement errors were small, and it outperformed all models except for a linear state-space model when measurement errors were large (the top three models are shown in Fig. 2B, see Fig. S1 for the full set). The state-space UDE outperformed the two most competitive models from the one-step ahead tests over a ten-year forecasting horizon, except for the Gaussian process model (gpEDM) when measurement errors were small and process noise was large (Fig. 2C). The state-space UDE model was also able to recover a chaotic attractor that closely matches the long-term behavior of the simulation model (Fig. 2D), demonstrating its ability to recover the qualitative behavior of the system in addition to making quantitative forecasts (Fig. 2E).

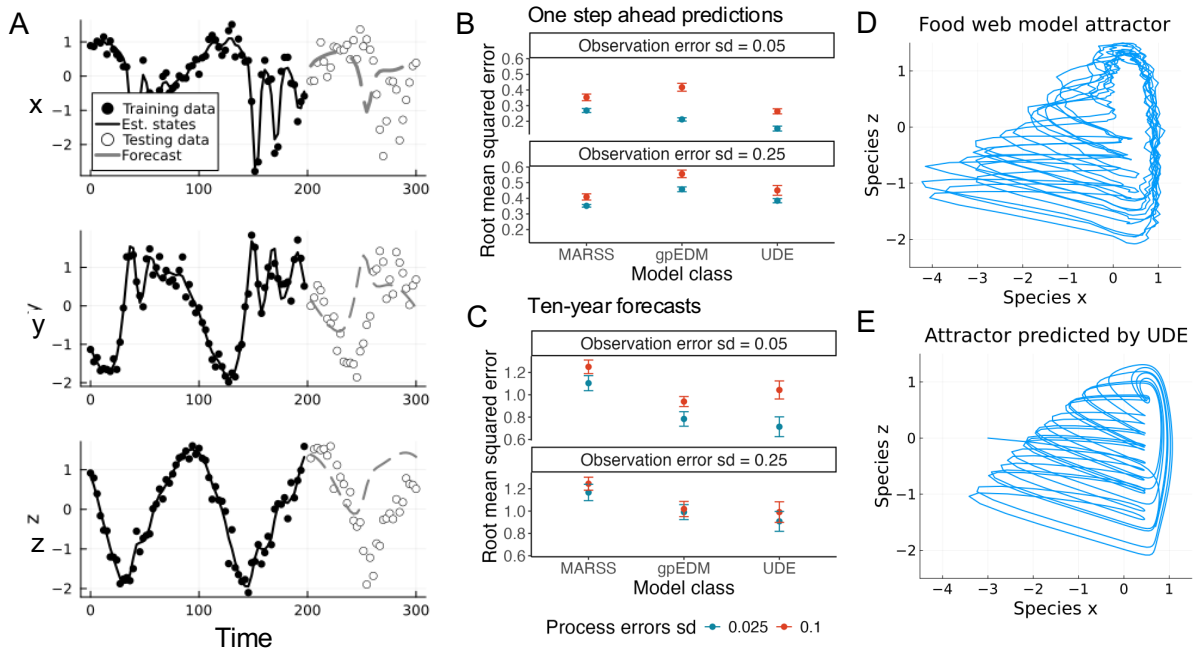

**Figure 2:** Performance of a state-space UDE model trained on simulated data from a three-species food chain. A) Time series of the training and testing data compared to the estimated unobserved states and forecasted abundances. B) The mean and standard error of the one-step ahead forecasting skill of the state-space UDE compared to an empirical dynamic model (gpEDM) and multivariate linear state-space model (MARSS) across 25 simulation tests. C) The mean and standard error of the forecasting skill of the state-space UDE, gpEDM, and MARSS models over a ten-year horizon across 25 simulation tests. D) Chaotic attractor of the three-species food chain model. E) Attractor predicted by the trained UDE.

*Recovering species interactions that cause transitions between alternative stable states*

We used a model of nearshore rocky reefs to test the ability of the state-space UDE framework to recover nonlinear species interactions that produce ecological tipping points and alternative stable states. In temperate latitudes, nearshore rocky reefs can switch between a kelp forest state dominated by macroalgae and an urchin barren state where sea urchins reduce kelp populations to very low densities (Estes & Duggins 1995; Konar & Estes 2003; Rogers-Bennett & Catton 2019). One mechanism that might cause these alternative states to persist is sea urchin behavior in response to food availability (Harrold & Reed 1985). When kelp are abundant, urchins exhibit cryptic behavior in which they hide from predators and feed on drift algae detached from the canopy or understory (Harrold & Reed 1985). In contrast, when kelp are sparse, urchins leave cryptic microhabitats to feed actively on live macroalgae (Fig. 3A). This change in behavior can cause urchin grazing to increase as kelp abundance declines (Fig. 3B), creating a feedback that amplifies variation in kelp abundance due to changes in abiotic conditions and causing the system to switch between kelp-dominated and urchin barren states (Fig. 3C).

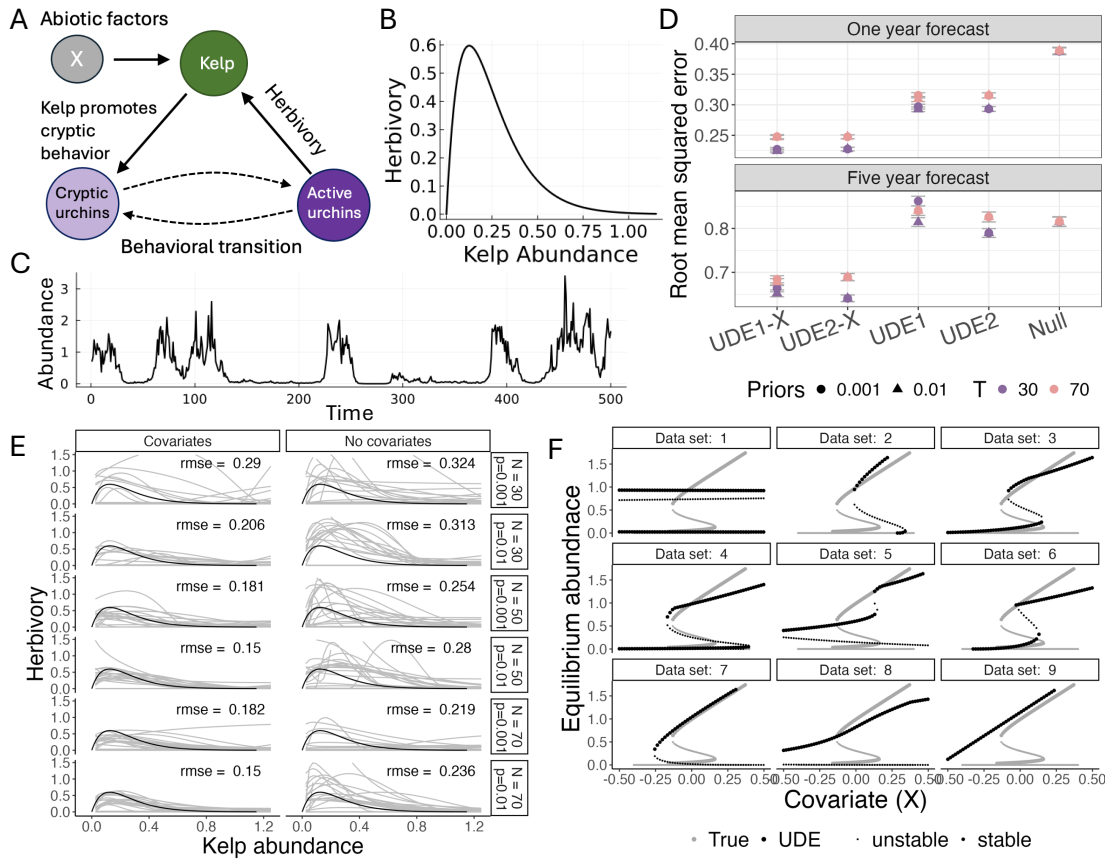

**Figure 3:** Estimates of nonlinear functions that produce ecological tipping points in kelp forests using state-space UDEs. A) Schematic diagram of the feedback between kelp abundance and sea urchin behavior. B) The nonlinear urchin grazing functional response to kelp abundance. C) Time series of kelp abundance illustrating the flickering behavior caused by the nonlinear urchin grazing function. D) The forecasting skill of parametric UDE models with and without covariates (UDE1-X, UDE1), nonparametric UDE models with and without covariates (UDE2-X, UDE2), and a null model over one- and five-year horizons when trained on time series of length T = 30 years and T = 70 years. The whiskers represent the standard error of the root mean square error (RMSE) estimates. E) The estimated relationship between kelp abundance and urchin herbivory from UDE1-X (left) and UDE1 (right) trained with different quantities of data, amount of prior information, and with and without covariates. F) Tipping points predicted by UDE 1-X trained on nine different simulated datasets each with a length 50 years.

We simulated time series of kelp abundance with a discrete-time model that describes changes in kelp abundance over time given the nonlinear urchin grazing functional response and autocorrelated abiotic effects on growth. We then fit a parametric UDE model (UDE1) to the simulated time series that mirrored the structure of the simulation model but used a neural network $NN(y_t; \boldsymbol{w}, \boldsymbol{b})$ to represent the unknown urchin grazing functional response. We first tested the ability of UDE1 to recover the urchin grazing response. We then tested its forecasting skill in comparison to a fully nonparametric state-space UDE (UDE2) as well as a random walk null model.

The parametric state-space UDE model (UDE1) was able to recover the characteristic nonlinearity in the urchin grazing functional response (Fig. 3E). Including abiotic conditions $\boldsymbol{X}_t$ in the models (UDE1-X, UDE2-X) and increasing the length of time series ($N$) improved the model estimates (Fig. 3D). Increasing the informativeness of the priors over the known parameters only helped when covariates were also included (Fig. 3E). In some simulations, the model predicted no relationship between kelp abundance and grazing. This was especially common with short time series (30 years) and is likely associated with datasets that span periods when the system was entirely in the kelp-dominated or urchin barren state, because the nonlinear urchin grazing function is not needed to describe a time series near a single equilibrium. Recovering the nonlinear urchin grazing response allows UDE-1 to detect alternative stable states and the ecological threshold separating them (Fig. 3F, Fig. S2), although the model was not able to detect the presence of a threshold from all simulated datasets.

Although the UDE models were effective at recovering the nonlinear relationships that determined the dynamics of the simulated data, they had more limited forecasting skill. The UDE models outperformed the null model over a one-year horizon, but the UDE models without covariates were outperformed by the null model over a five-year forecasting horizon (Fig. 3D). Adding known functional forms had little effect on the model's forecasting ability in this example. UDE1, which included the functional form of density dependence, and the fully non-parametric UDE2 had similar forecasting skills over both time horizons.

*Forecasting oscillating dynamics in fisheries data*

Commercial fisheries can exhibit boom and bust cycles characterized by periods of overexploitation that cause declines in the harvested stock followed by periods of low harvest and stock recovery (Fig. 4A,B, Wilen 2018). These so-called bioeconomic cycles are caused by delayed feedback between the abundance of the harvested population and the incentives of firms to invest in the fishery (Smith 1969). Fisheries management can smooth out these cycling dynamics by setting harvest limits that are linked to the abundance of the population (Free *et al.* 2022), creating more responsive feedback between the stock and harvest (Walters & Hilborn 1976).

We developed a state-space UDE model (UDE1) to quantify the feedback between the fish population abundance and harvest, and to forecast the dynamics of fisheries under different management scenarios. The UDE model describes changes in the population abundance $B$ and harvest rate $H$. We modeled the changes in abundance $B$ using the logistic model minus the harvest rate $H$. We modeled harvest as $H = UB/q$, where $U$ is fishing mortality and the factor $q$ ensures the units of harvest and abundance match. We modeled changes in fishing mortality with a neural network and derive a model for the resulting changes in harvest in the Methods. We compared the forecasting skill of this model to a fully nonparametric UDE model (UDE2).

We fit the UDE models to time series of harvest and abundance from two groundfish species, chilipepper rockfish (*Sebastes goodei*) and cowcod (*Sebastes levis*), sourced from the RAM Legacy Stock Assessment Database (2018). Both fisheries are from the Pacific coast of the United States and experienced major regulatory changes in the 1990s. To reflect this, we included an indicator variable $I_{t>1992}$ in the neural network that takes the value of zero pre-1992 or one post-1992 when access to the fisheries was restricted and other regulations were introduced (Pacific Fisheries Management Council 2019).

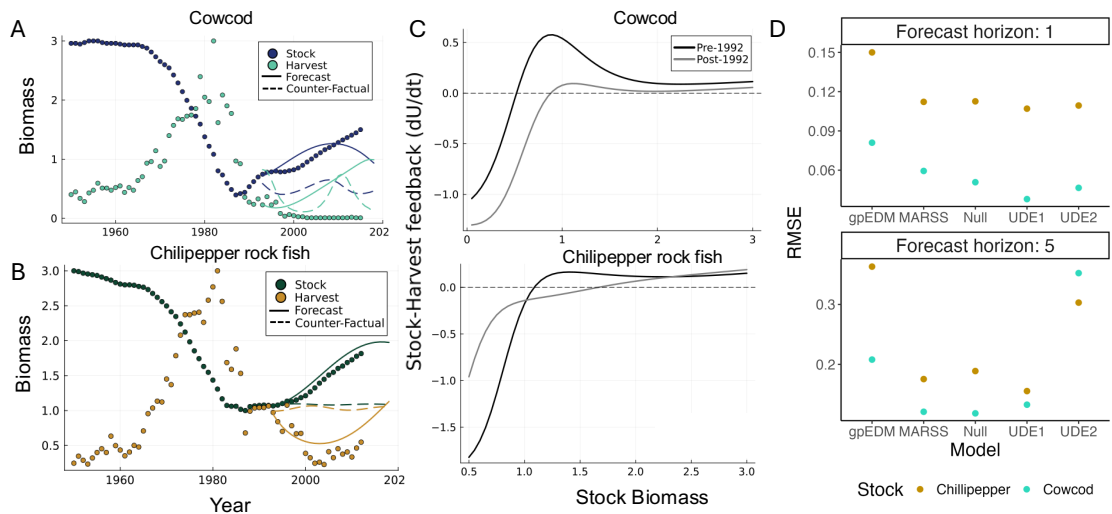

**Figure 4:** Forecasts of state-space UDE models for two groundfish fisheries. Time series of stock and harvest biomass with forecasts under the pre- and post-1992 management regimes for A) cowcod and B) chilipepper rockfish. C) Estimated relationships between the harvestable stock and the change in fishing effort over time for cowcod and chilipepper rockfish. D) Forecasting skill of UDE models with covariates (UDE-1, UDE-2), Gaussian process model (gpEDM), multivariate autoregressive state-space model (MARSS), and null model for one-year and five-year forecast horizons.

The parametric state-space UDE model (UDE1) predicted highly nonlinear relationships between the fish population biomass and changes in fishing mortality. Models for both groundfish predicted that fishing effort would increase relatively slowly when the stocks were abundant, and decline rapidly when the stocks were scarce (Fig. 4C). The stock-harvest feedback functions were qualitatively similar before and after the change in regulations in 1992, but the stock size where the change in fishing effort became negative was larger after 1992. This indicates that fishing effort became more responsive to fish population abundance after the change in management. Forecasts of the population abundances and harvests under the post-1992 and the counterfactual pre-1992 stock-harvest feedbacks suggest that the increases in abundance after 1992 can be attributed to the influence of new regulations on the stock-harvest feedback. The parametric state-space UDE model had the highest forecasting skill out of five models tested over a one-year horizon, and it performed similar to or slightly better than a linear state-space model (MARSS) and random walk null model over a five-year forecasting horizon (Fig. 4D). The fully nonparametric UDE (UDE2) performed relatively poorly in the five-year horizon case, indicating that adding parametric functions to a model can improve forecasts.

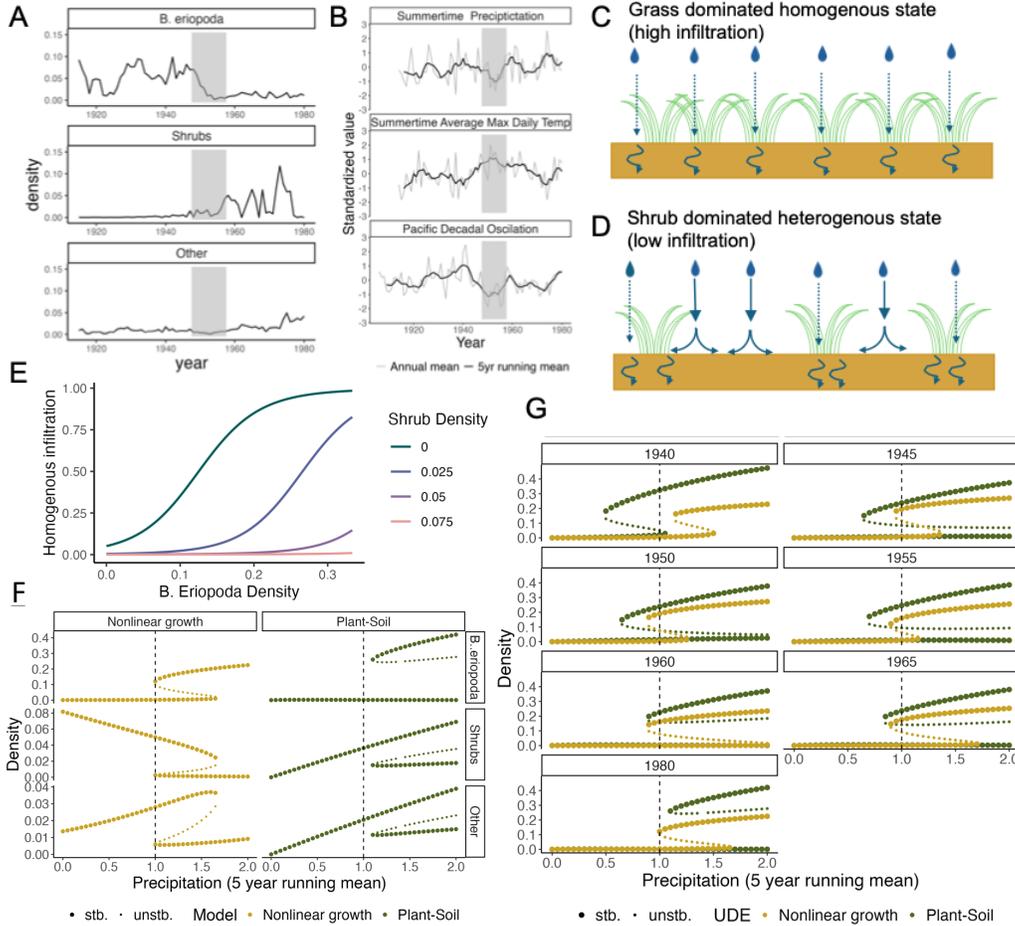

**Figure 5:** UDE models of alternative stable states in an arid rangeland. A) Time series of densities of the dominant grass species black grama (*Bouteloua eriopoda*), shrubs, and other vegetation species from sites with sandy soils in the Jornada Experimental Range in the Chihuahuan Desert, New Mexico. The period of community transition (grey shading) from a grass- to shrub-dominated state occurred around the year 1950. B) Time series of precipitation, temperature, and the Pacific Decadal Oscillation (PDO). The rangeland community transition highlighted in grey corresponds with decreased precipitation, increased temperatures and aridity, and a persistent negative PDO. Schematic diagrams of the rainfall infiltration of soil under the C) grass-dominated and D) shrub-dominated states. Under high grass densities, the velocity of falling precipitation and horizontal transport along the ground is reduced, resulting in relatively even soil infiltration and moisture levels. In the shrub-dominated state, rainfall landing on bare ground can flow more easily, concentrating in pools and producing heterogenous soil moisture. E) The estimated relationship between *B. eriopoda* and shrub densities and the rate of rainfall soil infiltration. F) Bifurcation diagram of the trained UDE model, showing the equilibrium abundance of each species group as a function of precipitation. G) Nonlinear growth and plant-soil feedback UDE model predictions of equilibrium grass density as a function of precipitation based on subsets of data starting in 1915 and ending in the year indicated by the panel title.

*Identifying ecological thresholds in an arid rangeland*

We tested the ability of state-space UDEs to recover ecological thresholds in an empirical dataset using long-term monitoring data from the Jornada Experimental Range (JER) in the Chihuahuan Desert of New Mexico, USA (Christensen *et al.* 2021, Fig. 5A). Plant communities in arid rangelands like the JER can exhibit rapid changes between states dominated by perennial grasses, woody forbs, and bare ground (Briske *et al.* 2005, Fig. 5C,D). Transitions between community states can be caused by the management of livestock grazing, abiotic factors like temperature and precipitation, and local feedbacks between local plant abundances and soil conditions (Briske *et al.* 2003; Schlesinger *et al.* 1990). For example, grasses can increase the ability of rainfall to infiltrate soils and reduce horizontal transport of water and nutrients. This results in a homogeneous distribution of soil moisture that promotes the grass-dominated community (Fig. 5C). However, if exogenous drivers such as heavy grazing or low precipitation cause grass densities to decline, then infiltration will decrease, resulting in greater horizontal transport and a heterogenous distribution of soil moisture that promotes a shrub-dominated state (Fig. 5D, D'Odorico *et al.* 2013; Schlesinger *et al.* 1990)

The JER dataset documents a regime shift from a grass-dominated state to a shrub-dominated state starting in the mid-1950s when drought conditions caused an abrupt decline in grass densities, followed by shrub encroachment over the second half of the twentieth century (Christensen *et al.* 2023, Fig. 5A,B). We developed two state-space UDE models of community composition in the JER to test their ability to recover the feedback mechanisms that produced the regime shift. The first model (plant-soil UDE) uses known functional forms to mechanistically represent the feedback between plant community composition and soil conditions hypothesized to cause regime changes in arid rangelands. The second (nonlinear-growth UDE) uses a neural network to estimate species interactions and known functions to capture competition for space. We compared these two models to a random walk null model.

The plant-soil UDE predicted a positive effect of grass density and a negative effect of shrubs on rainfall infiltration (Fig. 5E), matching the patterns reported in the literature (e.g., Schlesinger *et al.* 1990). This relationship stabilizes alternative grass-dominated and shrub-dominated states separated by a fold bifurcation at low precipitation (Fig. 5F). The nonlinear growth UDE also identified a fold bifurcation as a function of precipitation (Fig. 5F). However, these results were sensitive to the strength of regularizing (smoothing) priors over the neural network weights, with neither model predicting a regime shift when regularization was increased (Fig. S7).

We tested the ability of the UDEs to identify the fold bifurcation before the regime change in the 1950s by training the models on subsets of data leading up to and then immediately after the regime change. Both models were able to predict a fold bifurcation before the regime change in 1950. However, the plant-soil feedback model predicted the regime change would occur at much smaller values of precipitation when trained on datasets that did not include the 1950s drought (Fig. 5G). Regularizing priors over the neural network parameters also influenced the results of this analysis, causing neither model to predict a tipping point before the regime change (Fig. S8).

Although the two UDE models were able to recover the fold bifurcation consistent with the community time series, their performance on out-of-sample data was only slightly better than a simple null model (Table S5). Both UDEs outperformed the null model in leave-future-out cross-validation tests on a one-year forecasting horizon (plant-soil feedback UDE RMSE = 0.098, nonlinear growth UDE RMSE = 0.115, null model RMSE = 0.117), but performed equally well or worse than the null model over a five-year time horizon (Table S5). This is consistent with our results from the kelp-urchin example where models successful at inferring alternative regimes did not greatly increase forecasting skill over simple heuristics.

**Discussion**

We show that state-space UDEs can recover nonlinear interactions that produce complex ecosystem dynamics in the context of environmental stochasticity and noisy ecological data. As a result, this new class of models can represent a wide range of dynamic phenomena in ecology including chaos, regime shifts, and population cycles. Our findings from the Jornada Experimental Range dataset even suggest that state-space UDEs can identify alternative stable states and ecological thresholds before a regime shift occurs. Because of their ability to recover nonlinear dynamics from noisy data, state-space UDEs can be valuable both for forecasting ecosystem dynamics and as interpretive tools for inferring species interactions and feedback between organisms and their environment.

State-space UDEs perform well at these tasks by smoothing the dataset to account for observation errors and estimating the nonlinear functions that determine the system's dynamics while accounting for uncertainty. Accounting for imperfect measurements (observation errors) can reduce overfitting to training data by limiting the influence of noise from the observation process on the trained model. Accounting for uncertainty (process errors) in the model's predictions reduces the likelihood that this smoothing process removes real variation in the system's states that cannot be predicted from previous observations and covariates alone. This property is necessary to accurately represent data from systems that are inherently random (e.g., demographic stochasticity in finite populations) or systems that are influenced by unobserved exogenous factors.

UDEs may perform well at specific tasks because of their ability to combine prior information from known functional forms and parameters with functions learned from the data. This combination is necessary for tasks that involve inferring specific unknown functions (e.g., the feedback between rainfall and grass density). In these cases, the known functions may be required for the function represented by the neural network to have biologically meaningful interpretations. Including known functional forms can also be valuable for modeling systems with multiple dynamic regimes (e.g., alternative stable states). In the Jornada rangeland example, we found that a UDE model explicitly encoded with plant-soil feedbacks was able to predict the presence of alternative states before a regime shift occurred. The known functions used in this model explicitly encoded the feedback mechanisms stabilizing the two alternative states, which allowed the model to extrapolate beyond the dynamic regimes observed in the training data.

Despite their general usefulness for learning nonlinear dynamics in all our examples, the value of state-space UDEs for forecasting was context-dependent. The UDEs were usually the most skillful forecasting tool for the chaotic food chain model and the fisheries dataset, where the intrinsic nonlinear dynamics of the system produced large variability in the time series. However, in the kelp forest model and the Jornada rangeland dataset, the time series persisted near one of two alternative stable states. During these periods of relative stasis, simple models that explained one regime well were better at forecasting relative to more complex models that could capture both regimes.

Adding known functions to the state-space UDE models had mixed effects on forecasting skill in our examples. Additional known functional forms increased forecasting ability in the fisheries bioeconomic example, and reducing forecasting in the kelp forest and range land examples over a five year time horizon. This is consistent with prior studies on forecasting that show fully nonparametric models can outperform correct parametric models (Perretti *et al.* 2013), and in similar early experiments with standard UDE models (Arroyo-Esquivel *et al.* 2023). Taken together, these results suggest that adding parametric functional forms to UDEs can improve quantitative forecasts, but their value is case dependent. Our results suggest that adding known functional forms can improve the state-space UDE model's ability to recover qualitative features of the systems dynamics, which can be important for using other methods for anticipating large and rapid regime shifts, such as the use of early-warning signals based on critical slowing down (Boettiger *et al.* 2013; Scheffer *et al.* 2009).

Some advantages of state-space UDEs are shared by existing methods. Standard implementations of UDEs can recover a wide range of nonlinear dynamics, including species interactions from ecological datasets (Bonnaffé *et al.* 2021) along with several types of bifurcations such as the fold bifurcations we observed here (Vortmeyer-Kley *et al.* 2021). However, to our knowledge, the ability of UDEs to recover bifurcations from empirical ecological datasets has not previously been demonstrated. Furthermore, noisy or highly variable datasets can cause simple UDE training procedures to converge on suboptimal solutions (Turan & Jäschke 2022). Although these issues can be mitigated by modifying the loss function used to train the model (Turan & Jäschke 2022; Vortmeyer-Kley *et al.* 2021), state-space UDEs maintain these advances while explicitly accounting for uncertainty.

One unique feature of state-space UDEs is the variety of extensions that are possible by modifying the observation model. The distributional assumptions over the observation error can be changed to accommodate different data types. In place of a normal distribution, a Poisson distribution could be used for count data or a binomial distribution could be applied to frequency data. We restricted our analyses to observation models that used a one-to-one mapping between states and observations. However, observation models that project data into a lower-dimensional space may be valuable for modeling high-dimensional datasets (e.g., Zuur *et al.* 2003).

Altogether, state-space UDEs are a powerful new method for uncovering nonlinear ecosystem dynamics in the context of multiple forms of uncertainty common to ecological datasets. UDE models link what we know to be true from observational and experimental studies to flexible mathematical functions and then train over time series data from long-term environmental monitoring. As a result, they can encode mechanisms identified from empirical studies and estimate their influence on system dynamics in a quantitatively rigorous framework. By utilizing multiple sources of information in this way, state-space UDEs can be a valuable tool for synthesizing ecosystem data and, under the right conditions, forecasting ecological change.

## Methods

*Training state-space UDEs*
We trained the state-space UDE models by maximizing the log-likelihood function (equation 2) using the Adam (Kingma & Ba 2014) gradient descent algorithm implemented in the Optimizers.jl (Dixit & Rackauckas 2023) library in the Julia programming language (Bezanson *et al.* 2017). We computed the gradients of the loss function via automatic differentiation using Zygote.jl (Innes 2019). For continuous-time models, we numerically solved the integral in equation 1 using the Tsit-5 algorithm (Tsitouras 2011) implemented in DifferentialEquations.jl (Rackauckas & Nie 2017) and use DiffEqFlux.jl (Rackauckas *et al.* 2021) for compatibility between the ODE solvers and the automatic differentiation. We implemented this training procedure using the UniversalDiffEq.jl library (Buckner *et al.* 2024).

*Cross-validation*
We estimated the state-space UDE model performance on out-of-sample data using *k*-fold cross-validation and leave-future-out cross-validation. Leave-future-out cross-validation uses the beginning of the dataset to train the model and uses the final observations as testing data. The algorithm constructs multiple testing datasets by sequentially reducing the length of the training dataset, removing the final observation in the training dataset and adding it to the testing dataset. We calculated the model's performance by forecasting from the estimated value of the state variables at the final time point in the training data $\hat{u}_T$ over the full length of the testing dataset, which we varied in length to test the model's performance on different forecasting horizons. We quantified the accuracy of the forecasts using root mean square error (RMSE).

We implemented *k*-fold cross-validation by breaking the dataset up into *k* equal-sized segments of consecutive observations. Each segment is used as a testing dataset with the remaining blocks forming the training set. We calculated the model's performance predicting one time step ahead on the testing data. Each prediction step starts at the estimated value of the state variables at the preceding observation and is compared to the observation in the dataset. We calculated the estimates of the state variables in the testing dataset using a particle filter algorithm where the update step is determined by the trained state-space UDE model.

*Three-species food chain simulation tests*
We extended the Hastings and Powell (1991) three-species food chain model to include a source of stochasticity by adding a multiplicative white noise term with variance $\sigma_x^2$ to the basal resource yielding a system of three stochastic differential equations:

3) $\mathrm{d}x = x(1-x) - \frac{a_1 xy}{1+b_1 x} + \sigma_x x \mathrm{d}W$
4) $\mathrm{d}y = \frac{a_1 xy}{1+b_1 x} - \frac{a_2 yz}{1+b_2 y} - d_1 y + 0 \mathrm{d}W$
5) $\mathrm{d}z = \frac{a_2 yz}{1+b_2 y} - d_2 z + 0 \mathrm{d}W$.

We simulated time series with 100 observations sampling at an interval of 3.333 units of model time. We standardized the time series to have mean zero and variance one before adding normally distributed measurement error with mean zero and standard deviation $\sigma_\epsilon$. The remaining model parameters (Table S1) were chosen so that the model would exhibit chaotic dynamics in the absence of the white noise term. All models were trained on the first 75 observations and the last 25 were used to test the model's forecasting ability.

We simulated 25 datasets with four combinations of process and observation noise (100 total): high process noise $\sigma_x = 0.1$ and high observation noise $\sigma_\epsilon = 0.25$, low process noise $\sigma_x = 0.025$ and low observation noise $\sigma_\epsilon = 0.05$, and the two mixed cases. We compared the forecasting skill of the UDE model to a random walk null model, a multi-variate empirical dynamic model (mvEDM, Ye & Sugihara 2016), a Gaussian process empirical dynamic model (gpEDM, Munch & Rogers 2024), and a multivariate autoregressive state-space model (MARSS, Holmes *et al.* 2012). The full details of these alternative models are provided in supplemental appendix 1.2. We evaluated the one-step ahead forecasting ability of each model by predicting each state transition in the testing dataset. The null model and the two empirical dynamic models made predictions directly between the observed data points without accounting for measurement errors. The state-space UDE and MARSS models predicted the next observation from the estimated value of the state variables at the prior observation, which were estimated by a particle filter and Kalman filter algorithm, respectively. The ten-step forecasts were calculated by predicting the first ten observations in the testing set starting from the final time point of the training set.

*Kelp forest model simulation tests*

We simulated time series of kelp abundance $y_t$ using a discrete time model that accounted for density-dependent growth, urchin herbivory $h(y_t)$, observed abiotic factors $X_t$ and unobserved abiotic factors $v_t$. We modeled a density-dependent effect on growth using the Ricker model with additive effects for herbivory and abiotic conditions:

6) $y_{t+1} = y_t e^{r(1-y_t/K) - h(y_t) + X_t + v_t}$.

We selected a functional form for urchin functional response to kelp that increases at very low kelp densities up to a maximum value, and then declines at high and intermediate kelp densities to match the qualitative patterns of the kelp-urchin behavioral feedback:

7) $h(y_t) = \alpha y_t \exp(-\beta y_t)$.

The abiotic conditions were modeled as auto regressive processes (AR-1) with equal variance. The model parameters were selected to produce time series that regularly flickered between the kelp forest and urchin barren states.

The UDE models used the same structure as the simulation model (eq. 6), but we replaced the urchin grazing rate with a neural network and applied a log-transform:

8) $\log(y_{t+1}) = y_t + r\left(1 - \frac{y_t}{K}\right) - NN(\log(y_t))$.

We simulated 50 time series for each case in our simulation tests. We evaluated the predicted urchin functional responses by integrating the squared difference between the true and predicted functions over the range of biologically meaningful values, which we defined from zero kelp abundance to 1.5 times the kelp carrying capacity

9) $\text{score} = \sqrt{\int_0^{3K/2} (h(u) - NN(\log(u)))^2 du}.$

We calculated the forecasting skill of each model by predicting from the end of the training dataset over the first $n$ data points in the test set. We compared these forecasts to a null model and a fully nonparametric state-space UDE model described in detail in supplemental appendix 2.2.

*Fisheries UDE models*

We modeled changes in the biomass of the harvest population $B$ with a logistic growth model with growth rate $r$, carrying capacity $K$, and harvest $H$:

10) $\frac{dB}{dt} = rB\left(1 - \frac{B}{K}\right) - qH,$

where $q$ is used to ensure the units of harvest and stock biomass match. We modeled changes in the rate of harvest as a function of changes in the biomass of the population and changes in fishing mortality rate $U$, which is equal to the fraction of the stock harvested per unit time $U = qH/B$. We used a neural network to describe the changes in fishing mortality over time as a nonlinear function of the stock and an indicator variable $I_{t>1992}$ for the regulatory changes that began in 1992:

11) $\frac{dU}{dt} = U \times NN(B, I_{t>1992}).$

We found that multiplying the neural network output by the current fishing mortality rate, so that the neural network represents the proportion change in fishing mortality, produced better predictions in cross-validation tests. Finally, we derived a model for harvest using the product rule:

12) $\frac{dH}{dt} = \frac{dB}{dt}\frac{U}{q} + \frac{dU}{dt}\frac{B}{q} = \frac{dB}{dt}U + NN(B, I_{t>1992})H.$

We fit separate models to the chilipepper rockfish and cowcod abundance and harvest time series. For each species, we compared the forecasting ability of the parametric UDE model to a random walk null model, a linear MARSS model, a Gaussian process empirical dynamic model, and a second UDE model that used a single neural network to predict changes in harvest and abundance. Forecasting skill was evaluated using leave-future-out cross-validation with 20 testing datasets formed by reducing the training dataset length by one between each test.

*Jornada Experimental Range Data*

We analyzed long-term monitoring data from permanent quadrats in the Jornada Experimental Range, sourced from Christensen et al. (2021). The data include observations from 89 quadrats spanning 71 years with records of 126 species. We restricted our analysis to quadrats with high initial densities of the perennial grass *B. eriopoda*. These quadrats were selected by a cluster analysis (supplemental appendix 4.1). The remaining quadrats exhibited distinct dominant species groups, temporal dynamics, and soil textures (supplemental appendix 4.1). We chose to restrict our analysis to three species groups, *B. eriopoda*, shrubs which were primarily honey mesquite (*P. glandulosa*), and others, because species other than *B. eriopoda* and *P. glandulosa* occurred infrequently and in low abundances. A detailed explanation of the data preprocessing is provided in supplemental appendix 4.1.

*Jornada Experimental Range UDE models*

The state-space UDE models for the Jornada Experimental Range each tracked the abundances of the dominant perennial grass species *B. eriopoda* $X_g$, shrubs $X_s$, and other species $X_r$. The change in abundance of species $i$, $X_i$, included a term for growth into empty space $g_i(\mathbf{X}, P, T)$ that depended on the abundance of each species group $\mathbf{X}$, precipitation $P$, temperature $T$, and a constant mortality rate $m_i$. These effects are captured by a system of ODEs:

13) $\frac{dX_i}{dt} = g_i(\mathbf{X}, P, T)(A - X_g - X_s - X_r) - m_i X_i,$

where $A$ is the total area of space described by the model.

The two UDE models differ in how they construct the function $g_i(\mathbf{X}, P, T)$. The nonlinear growth UDE uses a neural network to capture species interactions and linear terms for precipitation and temperature. The final value is transformed by the softplus function $\log(1 + \exp(x))$ to ensure the growth rate always has a positive value:

14) $g_i(\mathbf{X}, P, T) = \text{softplus}(NN(\mathbf{X}) + \beta_{i,P} P + \beta_{i,T} T).$

The plant-soil feedback model explicitly incorporates the effect of plant community composition on soil moisture by adding two additional dynamic variables to the model: the quantity of homogenously-distributed soil moisture $M$ and the quantity of heterogeneously-distributed soil moisture $H$. The quantities of soil moisture were determined by the rate at which precipitation infiltrates the soil minus the rate of evaporation. We assume that water either infiltrates the soil directly and enters the homogenous pool $M$ or concentrates on the surface and enters the heterogenous pool $H$. We model the rate at which rainfall enters the homogenous pool using a neural network that takes the abundance of each species group as inputs and applies the logistic transform to the output to obtain a value between zero and one. Given these assumptions the levels of the soil moisture pools can be expressed with a system of ODEs:

15) $\frac{dM}{dt} = NN(\mathbf{X})P - (E_{M,0} + \xi_M T)M$

16) $\frac{dH}{dt} = (1 - NN(\mathbf{X}))P - (E_{H,0} + \xi_H T)H,$

where evaporation is a linear function of temperature with slope $\xi_j$ and intercept $E_{j,0}$. The growth rate of each species group is proportional to the sum of each soil moisture pool weighted by $r_{i,j}$, where $i$ indexes over species and $j$ indexes over the two soil moisture pools:

$$17)\ g_i(\mathbf{X}, P, T) = \left(r_{i,M} M + r_{i,H} H\right).$$

We assume that the dominant grass species *B. eriopoda* only benefits from the homogenous soil moisture and therefore fix $r_{g,H} = 0$. We chose to approximate the values of $M$ and $H$ with their equilibrium values. This is often called a fast-slow approximation and assumes the dynamics of soil moisture stocks are fast relative to the other state variables, allowing us to derive a model that only depends on the current values of the plant community composition $\mathbf{X}$, Temperature $T$, and precipitation $P$ (SI appendix 4.3). We fit both models using the five-year running averages of temperature and precipitation following a prior study in this system which found that smoothed climate variables were better predictors of *B. eriopoda* abundances than raw climate variables (Christensen et al. 2023). We compared the two UDE models to a random walk null model using both *k*-fold and leave-future-out cross-validation.

**Acknowledgements:**
This research was supported by the National Science Foundation awards #2233982 and #2233983 to JRW and LCM on Model Enabled Machine Learning for Predicting Ecosystem Regime Shifts. This paper is a product of the model enabled machine learning for ecology working group, which includes the authors of the paper, Cheyenne Jarman, Kunal Rathore, and Emerson Arehart, all of whom provided valuable contributions to the intellectual environment that led to this paper. We would also like to thank Chris Rackauckas for help working with Julia Scientific Machine learning tools and the Hawai'i Institute of Marine Biology for hosting a workshop where the ideas for this project were developed.

**Author Contributions:**
Conceptualization: JHB, JRW, JAE, ZDM, AG
Software*:* JHB, JAE, NF, ZDM
Formal Analysis: JHB
Writing-Original Draft*:* JHB, ZDM
Writing Review and Editing*:* ZDM, AG, NF, JRW, LCM
Funding Acquisition: JAE, LCM, JRW
Supervision: JRW, LCM
Project administration: JRW, LCM

# Appendices

**Supplemental Appendix 1:** Three-species food chain simulation tests

*Supplemental Appendix 1.1: Simulation model parameters and procedure and data processing*

We simulate datasets from the stochastic three-species food chain model (Hastings and Powell 1991) by approximating solutions to main text equations 3-5 using the SRIW1 algorithm implemented in DifferentialEquations.jl. We then log-transformed and standardized the time series of each variable sampled at intervals of length $\Delta t$ and added a normally distributed noise term to each time and state variable combination. This process yielded time series of observations $\hat{x}_t$, $\hat{y}_t$ and $\hat{z}_t$ corresponding to each population series

S1) $\quad \hat{x}_t = \dfrac{\log(x(t)) - \frac{\Delta t}{T}\sum_{\tau \in \{0:\Delta t:T\}} \log(x(\tau))}{\sqrt{\frac{\Delta t}{T}\sum_{\tau \in \{0:\Delta t:T\}}\left(\log(x(\tau)) - \frac{\Delta t}{T}\sum_{i \in \{0:\Delta t:T\}} \log(x(i))\right)^2}} + \epsilon_{x,t}$

S2) $\quad \hat{y}_t = \dfrac{\log(y(t)) - \frac{\Delta t}{T}\sum_{\tau \in \{0:\Delta t:T\}} \log(y(\tau))}{\sqrt{\frac{\Delta t}{T}\sum_{\tau \in \{0:\Delta t:T\}}\left(\log(y(\tau)) - \frac{\Delta t}{T}\sum_{i \in \{0:\Delta t:T\}} \log(y(i))\right)^2}} + \epsilon_{y,t}$

S3) $\quad \hat{z}_t = \dfrac{\log(z(t)) - \frac{\Delta t}{T}\sum_{\tau \in \{0:\Delta t:T\}} \log(z(\tau))}{\sqrt{\frac{\Delta t}{T}\sum_{\tau \in \{0:\Delta t:T\}}\left(\log(z(\tau)) - \frac{\Delta t}{T}\sum_{i \in \{0:\Delta t:T\}} \log(z(i))\right)^2}} + \epsilon_{z,t}$

S4) $\quad \epsilon_{j,t} \sim N(0, \sigma_\epsilon)$.

We chose the model parameters that would produce chaotic dynamics in the absence of stochasticity based on the findings of Hastings and Powell (1991). These parameter values are listed in Table S1. We experimented with different values of process and observation errors as indicated in the table.

**Table S1:** Food chain model parameters

| parameter | Interpretation | value |
|---|---|---|
| $a_1$ | Attack rate of species $y$ on species $x$ | 5.0 |
| $a_2$ | Attack rate of species $z$ on species $y$ | 0.1 |
| $b_1$ | Functional response curvature $x$ - $y$ | 2.75 |
| $b_2$ | Functional response curvature $y$ - $z$ | 2.0 |
| $d_1$ | Mortality rate species $y$ | 0.4 |
| $d_2$ | Mortality rate species $z$ | 0.01 |
| $\sigma_x$ | Standard deviation of white noise term added to species $x$ | (0.025, 0.1) |
| $\sigma_\epsilon$ | Observation errors | (0.05, 0.25) |
| $T$ | Time series length | 333.3 |
| $dt$ | Sampling interval | 3.333 |

*Supplemental Appendix 1.2: Alternative models for three-species food chain data*

Null model
We compared each UDE model to a null model that predicted zero change in the state variables over time. We implemented the null models using the same methods as the state-space UDEs but used the identify function as the process model $F(u) = u$. To ensure the results between the UDE and null models were directly comparable we used the same process and observation weights to fit the models.

Multivariate empirical dynamic model
We used the implementation of multiview embedding for empirical dynamic models (EDMs) in the rEDM package to build fully a non-parametric nonlinear multivariate time series model to compare with the state-space UDE model. Multiview EDMs use a model averaging approach to combine observations from multiple predictors into one forecast. These models require inputs for specifying the number of dimensions in the dataset $D = 3$, the dimensionality of the dynamical system that generated the data or the embedding dimension $E = 3$, and the number of models to include in the average forecast multiview $= 84$. We chose multivariate and embedding dimensions to match the known dimensions of the model used to simulate the dataset and chose the number of models to average as a tradeoff between computation time and accuracy. The rEDM package only supports one-step-ahead forecasts so we did not include these models in the 10-year forecast horizon tests.

Gaussian process empirical dynamic model
We used Gaussian process empirical dynamic models implemented in the gpEDM R package. The gpEDM package support univariate time series, but can capture more complex dynamics by incorporating time lags into the model through a process called time delay embedding. We constructed a forecast for the multivariate dataset by fitting gpEDMs to the time series of each species and then averaging the forecasting ability of the three models. We set the embedding dimension of the models $E = 3$ to match the known dimensions of the system and set the length of time lags $\tau = 0$.

Multivariate Autoregressive State-space Model
We compared the UDE models to linear state-space models implemented using the MARSS R package. The equations for the process and observation models are given below, where $x_t, y_t$ and $z_t$ are the state variables, $\hat{x}_t, \hat{y}_t$ and $\hat{z}_t$ are the observations, and the remaining parameters are estimated from the data.

Process model:

$$\text{S5) } \begin{bmatrix} x_{t+1} \\ y_{t+1} \\ z_{t+1} \end{bmatrix} = \begin{bmatrix} \alpha_{x,x} & \alpha_{x,y} & \alpha_{x,z} \\ \alpha_{y,x} & \alpha_{y,y} & \alpha_{y,z} \\ \alpha_{z,x} & \alpha_{z,y} & \alpha_{z,z} \end{bmatrix} \begin{bmatrix} x_t \\ y_t \\ z_t \end{bmatrix} + \begin{bmatrix} v_{t,x} \\ v_{t,y} \\ v_{t,z} \end{bmatrix}$$

$$\text{S6) } \begin{bmatrix} v_{t,x} \\ v_{t,y} \\ v_{t,z} \end{bmatrix} \sim MvNormal\left( \begin{bmatrix} 0 \\ 0 \\ 0 \end{bmatrix}, \begin{bmatrix} q_{x,x} & q_{x,y} & q_{x,z} \\ q_{x,y} & q_{y,y} & q_{y,z} \\ q_{x,z} & q_{y,z} & q_{z,z} \end{bmatrix} \right)$$

Observation model:

S7) $\begin{bmatrix} \hat{x}_t \\ \hat{y}_t \\ \hat{z}_t \end{bmatrix} = \begin{bmatrix} x_t \\ y_t \\ z_t \end{bmatrix} + \begin{bmatrix} \epsilon_{t,x} \\ \epsilon_{t,y} \\ \epsilon_{t,z} \end{bmatrix}$

S8) $\begin{bmatrix} \epsilon_{t,x} \\ \epsilon_{t,y} \\ \epsilon_{t,z} \end{bmatrix} \sim MvNormal\left(\begin{bmatrix} 0 \\ 0 \\ 0 \end{bmatrix}, \begin{bmatrix} r & 0 & 0 \\ 0 & r & 0 \\ 0 & 0 & r \end{bmatrix}\right)$

One step ahead forecasting skill

We compared the one-step ahead forecasting skill of each of the alternative models to the state-space UDEs. We compared the forecasting skill for the best three models in our one-step ahead tests over a ten year horizon.

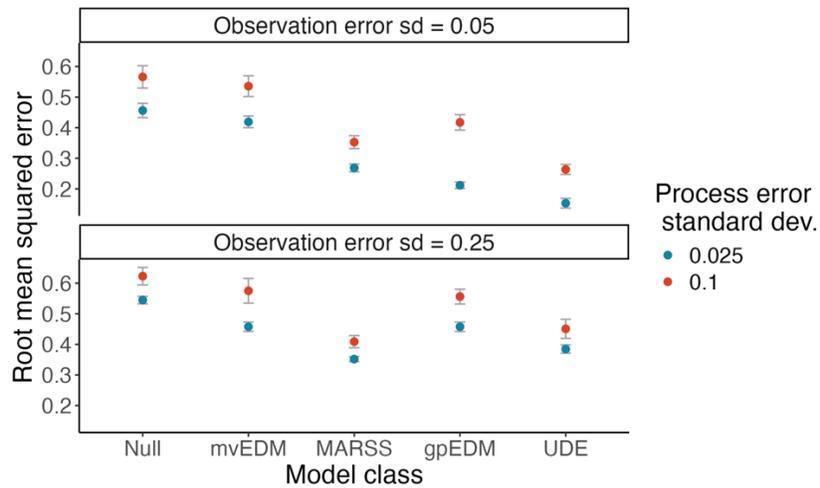

**Figure S1:** one-step ahead forecasting skill quantified by the root mean squared error of the UDE and four alternative models for different levels of observation (facets) and process errors (colors).

**Supplemental Appendix 2:** Kelp forest model simulation tests

*Supplemental Appendix 2.1*: Simulation model parameters

We chose parameters for the kelp forest model that produced time series that flickered between periods of high and low kelp abundance. We found these parameters experimentally by simulating time series until we found parameter combinations (Table S2) that resulted in kelp and urchin dominated states roughly 50% of the time. Each time series used to train the UDE models was generated by simulating the model for a burn-in period of 150 timesteps to reduce the influence of initial conditions on the datasets, then simulated for an additional 30, 50, or 70 years dependent on the number of timesteps required for the dataset. We then log-transformed the resulting time series and added normally-distributed observation errors at each time point with mean 0 and variance 0.25 before using them to train the models.

**Table S2:** Parameters for simulated kelp abundance time series

| PARAMETER | INTERPRETATION | VALUE |
|---|---|---|
| $r$ | Kelp per capita growth rate when rare | 0.5 |
| $K$ | Kelp carrying capacity | 1.0 |
| $a$ | Urchin grazing intensity | 13 |
| $b$ | Rate of decline in urchin grazing with increasing kelp abundance | 8 |
| $\sigma_X^2$ | Variance of observed environmental covariates | 0.055 |
| $\sigma_v^2$ | Variance of unobserved environmental conditions | 0.05 |
| $\rho$ | Autocorrelation of environmental conditions | 0.9 |
| $\sigma_\epsilon^2$ | Variance of observation errors | 0.125 |

*Supplemental Appendix 2.2*: Alternative model details

Null model
We benchmarked the kelp forest UDE model's forecasting accuracy by comparing it to a null model that predicts no change in the state variables between observations. We fit this model using the state-space modeling framework used for the UDE models but with the identity function for the process model $F(\boldsymbol{u}) = \boldsymbol{u}$. We used the same process and observation weights to estimate the unobserved state of the null model used to train the UDE models.

Nonparametric model
We also compared the kelp forest UDE model that included known functional forms to a fully nonparametric UDE model that used a neural network to learn the process model:

S9) $F(\boldsymbol{u}, X) = NN(\boldsymbol{u}, X)$.

*Supplemental Appendix 2.3:* Bifurcation diagrams

We tested the ability of the UDE to recover ecological tipping points in the simulated time series by fitting the models to 150 simulated datasets: 50 each of length 30, 50, and 70 observations, respectively. For each dataset, we visually compared the estimated bifurcation diagram to the true bifurcation diagram of the simulation model (Fig. S2).

**Supplemental Appendix 3:** *Forecasting oscillating dynamics in fisheries data*

*Supplemental Appendix 3.1:* Data processing

We obtained time series of historical catch and abundance for chilipepper rockfish (*Sebastes goodei*) and cowcod (*Sebastes levis*) from the RAM Legacy Stock Assessment Database (https://www.ramlegacy.org/). We rescaled both datasets to span from zero to three by dividing each time series by the maximum value and multiplying by three. Transforming the data in this way ensures that each time series is weighted equally in the loss function. We calculated a coefficient $q = \max(B_t) / \max(H_t)$ to ensure that the units of harvest and biomass matched for the interaction terms included in the models.

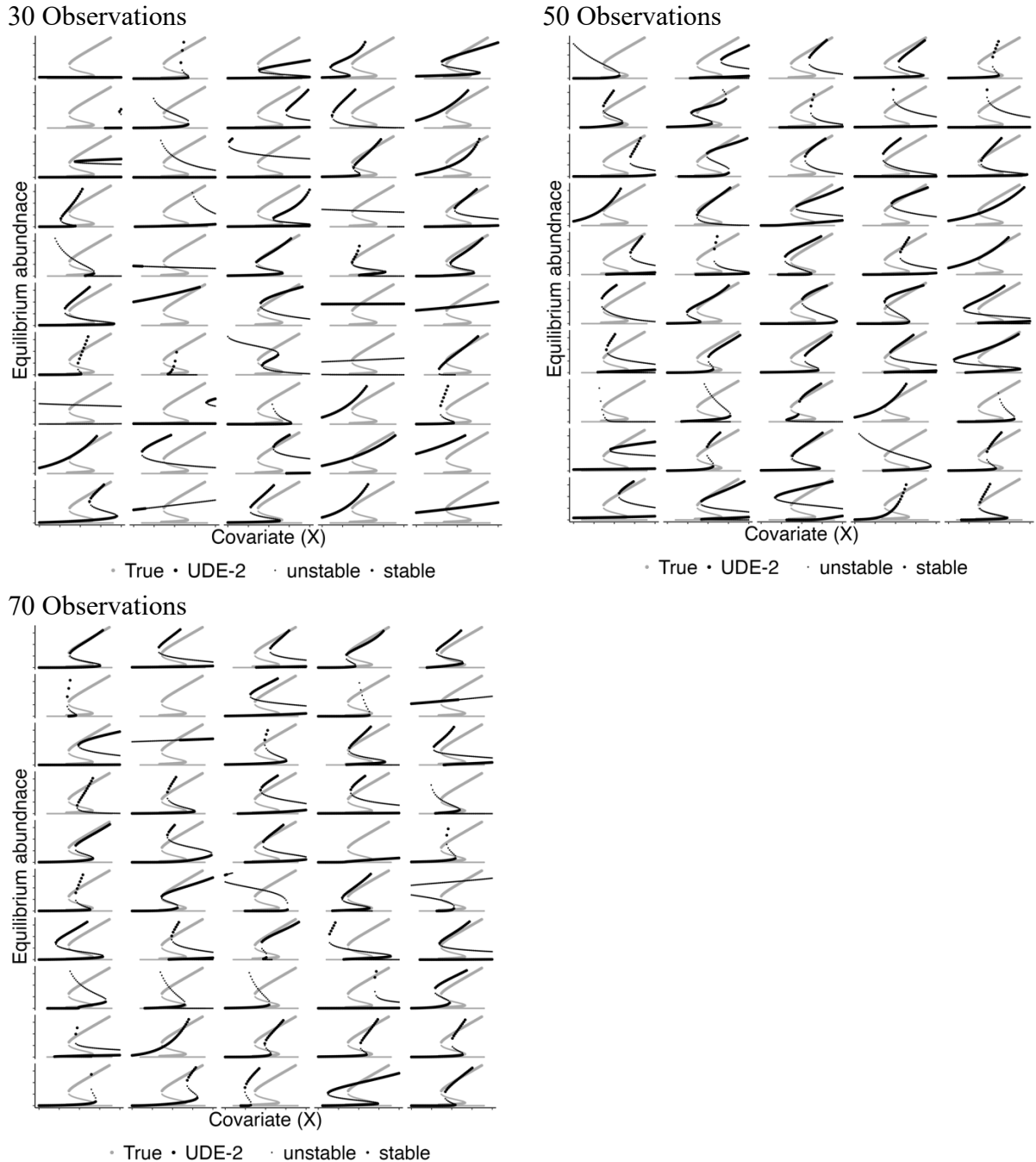

**Figure S2:** Bifurcation diagrams from the kelp forest simulation model (grey) and the UDE models (black).

*Supplemental Appendix 3.2:* Bioeconomic model derivation

Our dataset included observations of population biomass $B$ and harvest $H$. However, it makes more mechanistic sense to describe changes in population biomass $B$ and fishing mortality $U$. To this end, we defined an ODE model in terms of changes in biomass and fishing mortality and then used the relationship between fishing mortality harvest and biomass to derive a model for harvest. We assume harvest is equal to the product of the fishing mortality rate and biomass

S10) $H = UB/q$.

Therefore, the rate of change in harvest over time can be expressed using the product rule

S11) $\frac{dH}{dt} = \frac{dU}{dt}\frac{B}{q} + \frac{dB}{dt}\frac{U}{q}$.

Substituting main text equations ten and eleven into equation S11 and simplifying yields the final model

S12) $\frac{dH}{dt} = \left(U \times NN(B, I_{t>1992})\right)\frac{B}{q} + \left(rB\left(1 - \frac{B}{K}\right) - qH\right)\frac{U}{q}$,

$= \left(\frac{qH}{B} \times NN(B, I_{t>1992})\right)\frac{B}{q} + \left(rB\left(1 - \frac{B}{K}\right) - qH\right)\frac{H}{B}$,

$= NN(B, I_{t>1992})H + \left(rH\left(1 - \frac{B}{K}\right) - \frac{qH^2}{B}\right)$.

*Supplemental Appendix 3.3:* Alternative model details

Null model
We benchmarked the UDE model's forecasting accuracy by comparing it to a null model that predicts no change in the state variables between observations. We fit this model using the state-space modeling framework used for the UDE models but with the identity function for the process model $F(\boldsymbol{u}) = \boldsymbol{u}$. We used the same process and observation weights to estimate the unobserved state of the null model used to train the UDE models.

Gaussian process empirical dynamic model
We used the same procedure to construct Gaussian process EDM models for the fisheries time series as we used for the three-species food chain example, except we set the embedding dimension $E = 5$, because the true embedding dimension is unknown and gpEDMs use automatic relevance determination to prevent overfitting if too many time lags are included.

Multivariate Autoregressive State-space Model
We compared the UDE models to linear state-space models implemented using the MARSS R package. The equations for the process and observation models are given below, where $B_t$ and $H_t$ are the state variables, $\hat{B}_t$ and $\hat{H}_t$ are the observations, and the remaining parameters are estimated from the data.

Process model:

S15) $\begin{bmatrix} B_{t+1} \\ H_{t+1} \end{bmatrix} = \begin{bmatrix} \alpha_{B,B} & \alpha_{B,H} \\ \alpha_{H,B} & \alpha_{H,H} \end{bmatrix} \begin{bmatrix} B_t \\ H_t \end{bmatrix} + \begin{bmatrix} b_B \\ b_H \end{bmatrix} + \begin{bmatrix} v_{t,B} \\ v_{t,H} \end{bmatrix}$

S16) $\begin{bmatrix} v_{t,B} \\ v_{t,H} \end{bmatrix} \sim MvNormal\left(\begin{bmatrix} 0 \\ 0 \end{bmatrix}, \begin{bmatrix} q_{B,B} & q_{B,H} \\ q_{B,H} & q_{H,H} \end{bmatrix}\right)$

Observation model:

S17) $\begin{bmatrix} \hat{B}_t \\ \hat{H}_t \end{bmatrix} = \begin{bmatrix} B_t \\ H_t \end{bmatrix} + \begin{bmatrix} \epsilon_{t,B} \\ \epsilon_{t,H} \end{bmatrix}$

S18) $\begin{bmatrix} \epsilon_{t,B} \\ \epsilon_{t,H} \end{bmatrix} \sim MvNormal \left( \begin{bmatrix} 0 \\ 0 \end{bmatrix}, \begin{bmatrix} r & 0 \\ 0 & r \end{bmatrix} \right)$

**Supplemental Appendix 4:** Jornada rangeland example

*Supplemental Appendix 4.1: Data selection*

The Jornada Experimental Range long-term quadrat dataset spans several distinct plant communities, which were largely determined by soil characteristics at the quadrat site (Fig. S3). We categorized the quadrats by first projecting the community composition data into a lower-dimensional space using non-metric multidimensional scaling (NMDS) with the Bray-Curtis dissimilarity metric. The resulting NMDS fit was a good representation of data with a stress of 0.138. We grouped the quadrats based on their NMDS scores using the *k*-means clustering algorithm (Fig. S3B). Both NMDS scores and the *k*-means clusters were predicted by soil texture characteristics (Fig. S3C), with the most important predictors being the percents of sand, silt, clay and very find sand (Table S3). Community composition in clusters 4 and 5 (more negative in NMDS dimension 1) was associated with soils that had coarser textures (higher percent sand, lower percent clay, silt, and very fine sand). These quadrats tended to have high abundances of Black grama (*Bouteloua eriopoda*) at the beginning of the time series and were dominated by honey mesquite (*Prosopis glandulosa*) and other grass species after the regime shift in the 1950s (Fig. S3A). We chose to focus our analysis on this group of quadrats, which corresponded to *k*-means clusters 3, 4, and 5. These clusters were primarily differentiated by the abundances of honey mesquite and mesa dropseed (*Sporobolus flexuosus*) (Fig. S4, PRGLS2 and SPFL2).

**Table S3:** Loadings of the soil texture variables with the first two principal components. The values are the associated correlation coefficients (*r*), ranging from -1 to 1. The first principal component explained 48% of the variance, and the second principal component explained 29% of the variance.

| Variable | PC1 rotation | PC2 rotation |
|---|---|---|
| Percent sand | -0.48 | -0.11 |
| Percent silt | 0.48 | 0.12 |
| Percent clay | 0.44 | 0.09 |
| Percent very fine sand | 0.42 | -0.02 |
| Percent fine sand | -0.015 | -0.60 |
| Percent medium sand | -0.38 | 0.15 |
| Percent coarse sand | -0.16 | 0.59 |
| Percent very coarse sand | 0.01 | 0.48 |

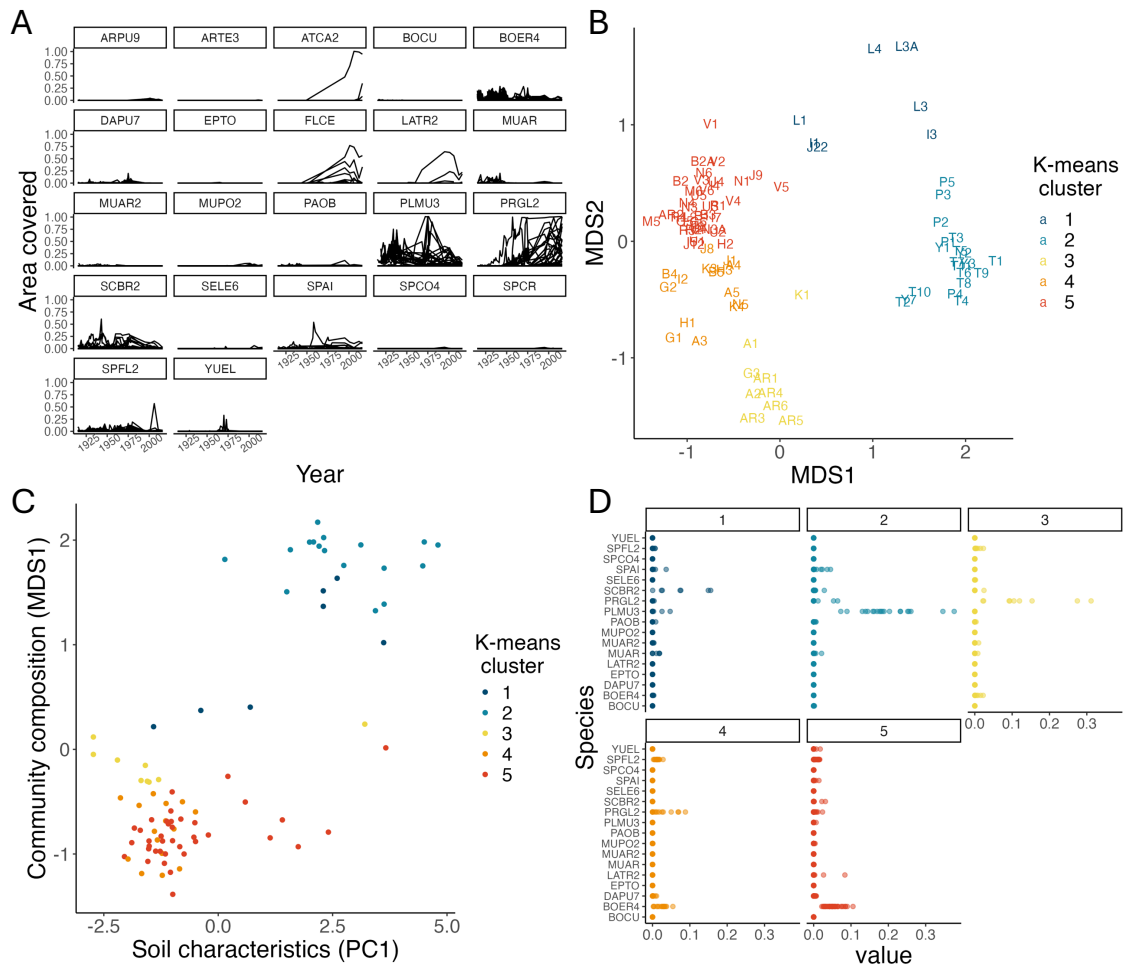

**Figure S3:** Jornada range plant community composition quadrat data. A) Raw time series data of the density of each species (facets) in each quadrat (lines). B) An NMDS plot of the average community composition in each quadrat from 1915 to 2016. C) Correlation between community composition (NMDS dimension 1) and soil characteristics (principal component 1). D) Average species composition in each quadrat, grouped by k-means algorithm into five clusters. See Table S4 for definitions of the species abbreviations.

**Table S4:** Species abbreviations, growth form, scientific and common names.

| Abbreviation | Genus | Species | Form | Common Name |
|---|---|---|---|---|
| SCBR2 | *Scleropogon* | *brevifolius* | GRASS | burrograss |
| PLMU3 | *Pleuraphis* | *mutica* | GRASS | Tobosagrass |
| BOER4 | *Bouteloua* | *eriopoda* | GRASS | black grama |
| SPFL2 | *Sporobolus* | *flexuosus* | GRASS | mesa dropseed |
| MUAR | *Muhlenbergia* | *arenacea* | GRASS | ear muhly |
| SPAI | *Sporobolus* | *airoides* | GRASS | alkali sacaton |
| DAPU7 | *Dasyochloa* | *pulchella* | GRASS | low woollygrass |
| PRGL2 | *Prosopis* | *glandulosa* | SHRUB | honey mesquite |
| YUEL | *Yucca* | *elata* | SHRUB | soaptree yucca |
| LATR2 | *Larrea* | *tridentata* | SHRUB | creosote bush |

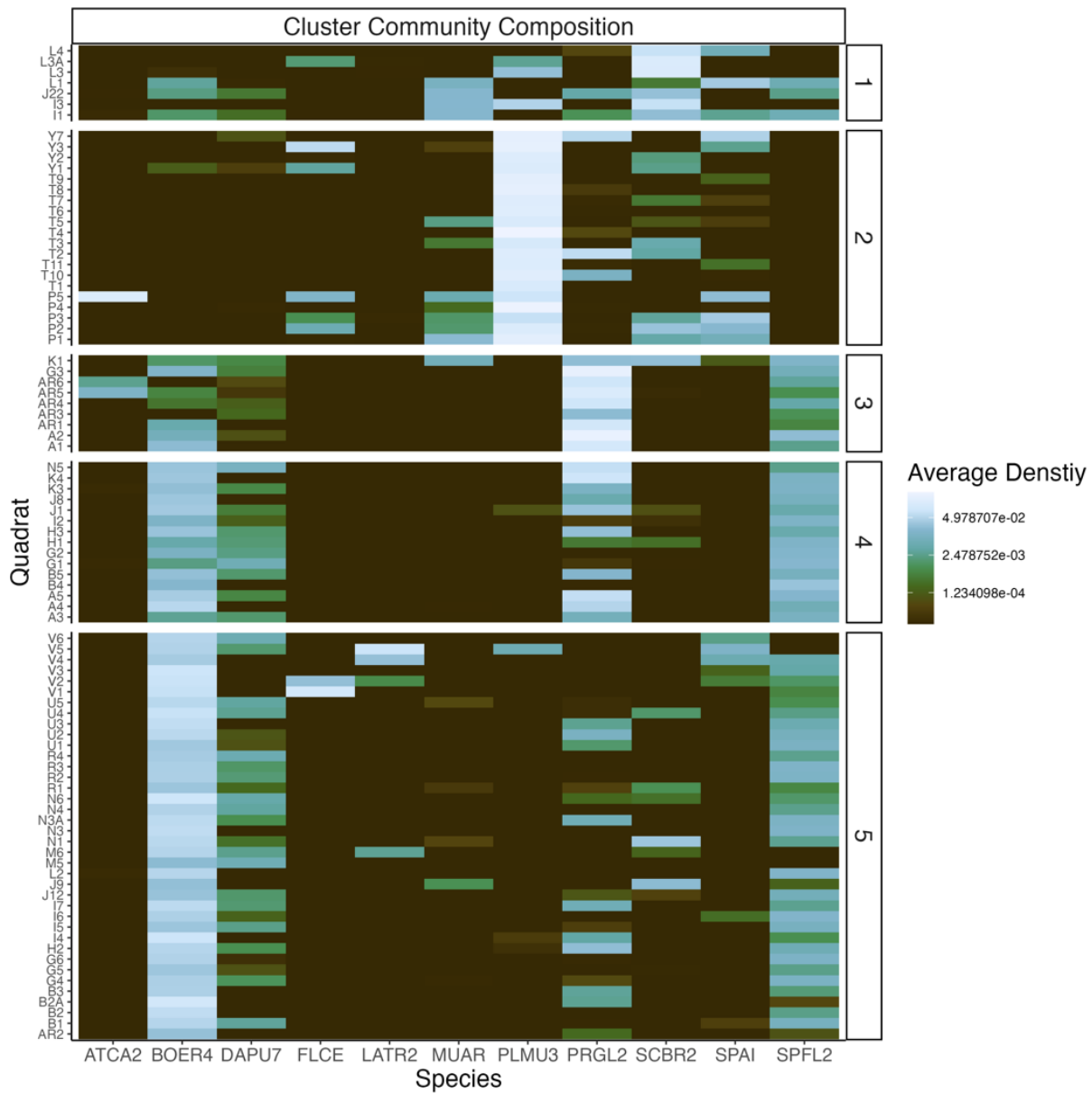

**Figure S4:** Average densities of the 11 most common species in each quadrat over the study period. The quadrats are grouped into the clusters identified by applying the k-means algorithm to NMDS based on community composition data.

We trained the UDE models on the average abundances of species within three species groups across all quadrats sorted into clusters 3, 4, and 5 of our community composition analysis. Not every quadrat was observed every year. This caused the annual average species abundances to vary over time, because quadrats with high abundances of certain species were not included in the average. This effect was particularly consequential for the shrub time series. Shrub-dominated communities are inherently patchy, and as a consequence, some quadrats were almost entirely covered by shrubs while others were almost entirely bare ground. The combination of this spatial pattern and inconsistent sampling appears to cause a high degree of variability in the shrub time series. To reduce the variability caused by the sampling procedure, we calculated the average abundance of each species group in each quadrat and used a linear interpolation to fill missing values in years when a quadrat was not sampled. We found this procedure had little or no effect on the *B. eriopoda* and other species time series but smoothed the shrub time series considerably (Fig. S5).

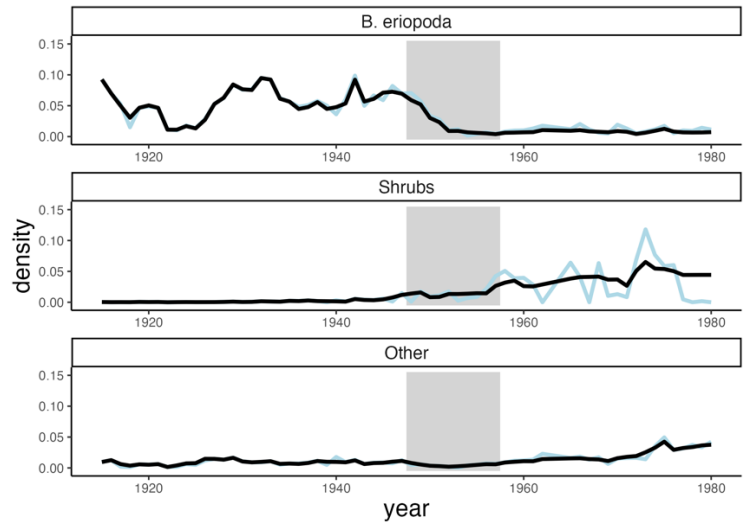

**Figure S5:** Comparison of annual plant densities averaged over monitored quadrats (blue line) to annual plant densities averaged over all quadrats using a linear interpolation to impute missing quadrat and year combinations (black line). The gray shading is the period of severe drought.

The shrub group was primarily composed of *Prosopis glandulosa*, but *Yucca elate* also appeared in low abundance (Fig. S6). The other species group was primarily comprised of three grasses: *Sporobolus flexuosus*, *Dasyochloa pulchella* and *Scleropogon brevifolius* (Fig. S6). Christensen et al. (2023) categorized these three species as transient grasses, which are often present in the Jornada rangeland, but rarely comprise large fractions of the plant community in a given area.

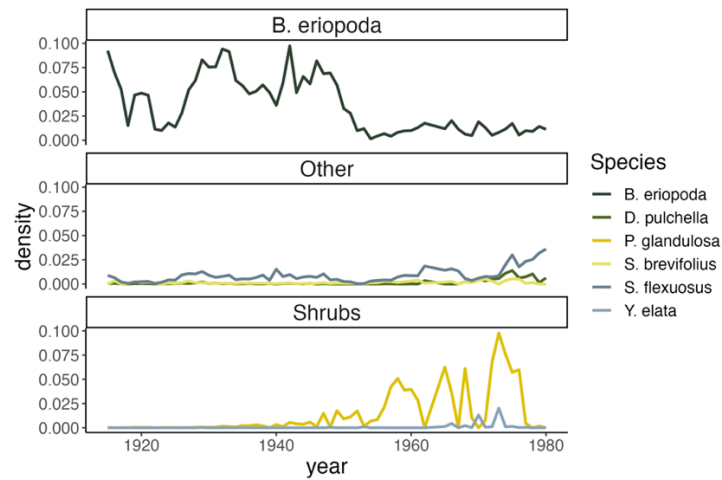

**Figure S6:** Annual densities of each species within the species groups.

*Supplemental Appendix 4.2: Derivation of plant – soil feedback model*

The model of plant soil feedbacks for the Jornada Experimental Range dataset included two state variables that represented the quantity and distribution of soil water content: a homogenously distributed stock $M$ and a heterogeneously distributed stock $H$. The dynamics of each stock are governed by a differential equation that depends on the composition of the plant community $\mathbf{X}$, precipitation $P(t)$, and temperature $T(t)$

S19) $\frac{dM}{dt} = f(\mathbf{X})P(t) - M(a_M + b_M T(t))$

S20) $\frac{dH}{dt} = (1 - f(\mathbf{X}))P(t) - H(a_H + b_H T(t))$.

These variables are not observed directly in our dataset, and the state-space UDE framework does not currently support unobserved dynamic variables. Because of this limitation, we chose to approximate the value of these variables by assuming the dynamics of the soil water pools is fast compared to changes in temperature, precipitation, and community composition. Although we cannot confirm the validity of this assumption without data on soil moisture levels, it allows us to capture the qualitative features of plant soil feedback without adding new state variables. Using this assumption, we set $M^*(t)$ and $H^*(t)$ at their equilibrium values, which we found by setting the right-hand side to zero and solving for $M$ and $H$, respectively. This yields approximate values of $M$ and $P$ that are functions of the current plant community composition, precipitation, and temperature:

S21) $M^*(t) = \frac{f(\mathbf{X}(t))P(t)}{(a_M + b_M T(t))}$

S22) $P^*(t) = \frac{(1 - f(\mathbf{X}(t)))P(t)}{(a_H + b_H T(t))}$.

*Supplemental Appendix 4.3: Alternative model details*

We benchmarked the UDE model's forecasting accuracy by comparing it to a null model that predicts no change in the state variables between observations. We fit this model using the same state-space modeling framework used for the UDE models, but we with the identity function for the process model $F(\mathbf{u}) = \mathbf{u}$. We used the same process and observation weights to estimate the unobserved state of the null model used to train the UDE models.

*Supplemental Appendix 4.4: Model comparison results*

We compared the two UDE models to the null model using leave-future-out cross-validation over one- and five-year time horizons (Table S5). We found the two UDE models outperformed the null model over the one-year time horizon and the nonlinear growth model performed almost equally to the null model over a five-year time horizon. The performance of the two UDE models depended on the strength of regularizing priors used on the neural network weights. More regularization (regularization = 3.5) generally improved the predictive skill of the plant-soil feedback model although the effect was small. The nonlinear growth model performed better in the low regularization case (regularization = 4.5).

**Table S5:** The forecasting skill (root mean squared error) of the UDE and null models over one- and five-year horizons.

| Model | Regularization | Horizon | total | *B. eriopoda* | Shrubs | Other |
|---|---|---|---|---|---|---|
| Plant-soil | 3.5 | 1 | 0.098 | 0.124 | 0.100 | 0.058 |
| Plant-soil | 4.5 | 1 | 0.103 | 0.136 | 0.100 | 0.058 |
| nonlinear | 3.5 | 1 | 0.117 | 0.165 | 0.102 | 0.057 |
| nonlinear | 4.5 | 1 | 0.115 | 0.162 | 0.101 | 0.057 |
| Null | 0 | 1 | 0.117 | 0.170 | 0.096 | 0.056 |
| Plant-soil | 3.5 | 5 | 0.277 | 0.401 | 0.233 | 0.129 |
| Plant-soil | 4.5 | 5 | 0.299 | 0.446 | 0.228 | 0.135 |
| nonlinear | 3.5 | 5 | 0.264 | 0.389 | 0.212 | 0.111 |
| nonlinear | 4.5 | 5 | 0.197 | 0.290 | 0.150 | 0.101 |
| Null | 0 | 5 | 0.197 | 0.307 | 0.117 | 0.094 |

*Supplemental Appendix 4.5: effect of regularization on UDE model predictions*

We chose to present the weak regularization cases (regularization = 4.5) because they captured more of the nonlinear species interactions, we hypothesized may be present in the data sets and either improved the model's performance on out of sample data in the nonlinear growth case or only caused a small decrease in the plant-soil feedback case. However, regularization did have a large effect on the dynamics predicted by the UDE models, specifically the UDE models with higher regularization generally predicted smoother relationships between precipitation and community composition (Figures S7 and S8).

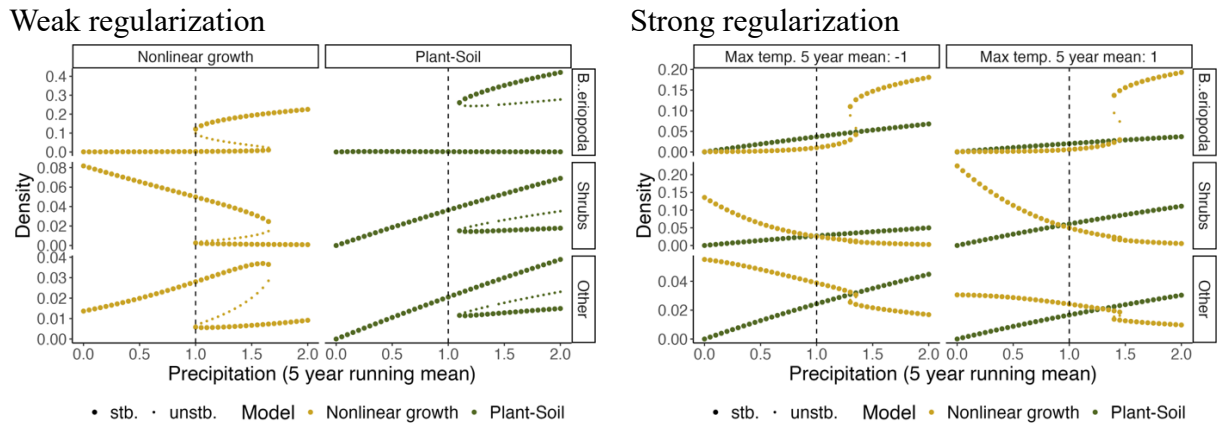

**Figure S7:** Bifurcation diagrams showing the relationships between precipitation and equilibrium plant densities predicted by the UDE models for two levels of regularizing priors.

Weak regularization · Strong regularization

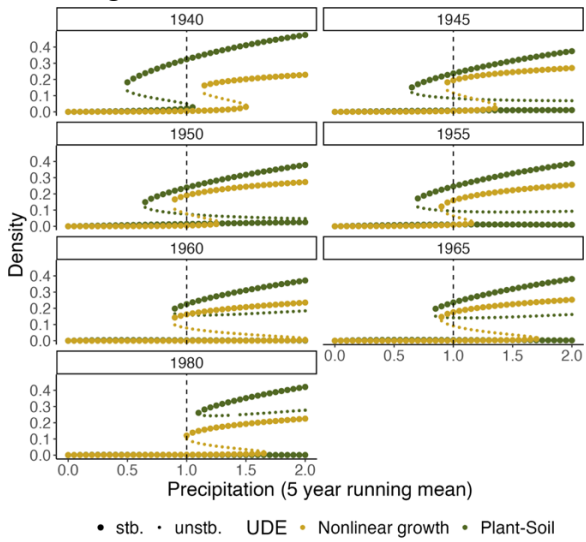
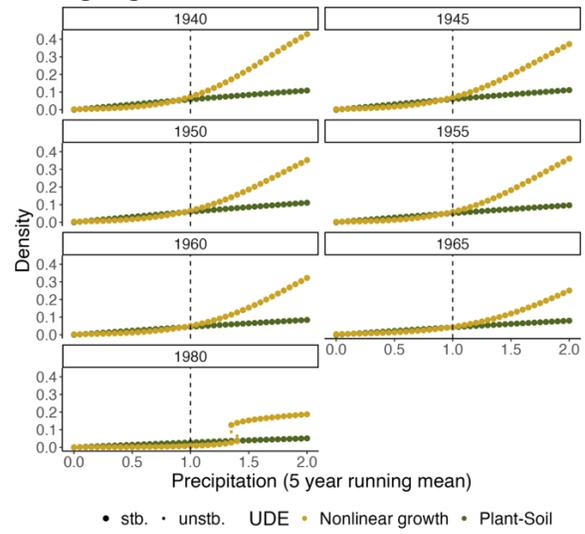

**Figure S8:** The predicted equilibrium B. Eriopoda density as a function of precipitation based on subsets of the data starting in 1915 and ending in the year indicated by the panel title, for the weak and strong regularizing priors